\documentclass{aa}  
\newcommand{\kms}{km~s$^{-1}$}
\newcommand{\sbr}[1]{\left[ #1 \right]} % square bracket
\newcommand{\rev}[1]{{#1}}

\usepackage{graphicx}
\usepackage{txfonts}
\usepackage{hyperref}
\begin{document} 

   \title{Probing gaseous halos of galaxies with radio jets}

   %\subtitle{I. Overviewing the $\kappa$-mechanism}

   \author{Martin G. H. Krause
          \inst{1}\fnmsep\thanks{\email{M.G.H.Krause@herts.ac.uk}}
          \and
          Martin J. Hardcastle\inst{1}%\fnmsep\thanks{Just to show the usage
          %of the elements in the author field}
	\and 
	Stanislav S. Shabala\inst{2}
          }

   \institute{Centre for Astrophysics Research, 
	School of Physics, Astronomy and Mathematics, University of Hertfordshire, 
	College Lane, \\ \:\,Hatfield, Hertfordshire AL10 9AB, UK\\
         \and
             School of Natural Sciences, Private Bag 37, University of Tasmania, Hobart, TAS, 7001, Australia\\
             }

   %\date{Received April 1, 2019; accepted ...}

% \abstract{}{}{}{}{} 
% 5 {} token are mandatory
 
  \abstract
  % context heading (optional)
  % {} leave it empty if necessary  
   {Gaseous halos play a key role for understanding inflow, feedback and the overall baryon budget in galaxies.
   Literature models predict transitions of the state of the gaseous halo between cold and hot accretion, winds, 
   fountains and 
   hydrostatic halos at certain galaxy masses. Since luminosities of radio AGN are sensitive
   to halo densities, any significant transition would be expected to show up in the radio luminosities of large samples of galaxies.
   The  Low Frequency Array (LOFAR) Two Metre Sky Survey (LoTSS) has indeed identified a galaxy stellar mass scale, $10^{11} M_\odot$,
   above which the radio luminosities increase disproportionately.}
  % aims heading (mandatory)
   {%The luminosity of radio-AGN depends on the
   %halo properties, but the distribution of jet powers is uncertain, so that their use for halo diagnostics is so far limited.
   Here, we investigate, if radio luminosities of galaxies, especially the marked rise 
   at galaxy masses around $10^{11}M_\odot$,
   can be explained with standard assumptions on jet powers, scaling between black hole-mass and galaxy mass
   and gaseous halos.}
  % methods heading (mandatory)
   {We developed models for the radio luminosity of radio AGN in halos under infall, galactic wind and hydrostatic
   conditions based on observational data and theoretical constraints, and compared it to LoTSS
   data for a large sample of galaxies in the mass range
   between $10^{8.5}M_\odot$ and $10^{12}M_\odot$.}
  % results heading (mandatory)
   {Assuming the same characteristic upper limit to jet powers as is known from high galaxy masses to hold at all masses,
   we find that the maximum radio luminosities for the hydrostatic gas halos fit well with the upper envelope of the
   distribution of the LOFAR data. The marked rise in radio luminosity at $10^{11} M_\odot$ is matched in our model,
   and is related to significant change in halo gas density around this galaxy mass, which is a consequence
   of the lower cooling rates at higher virial temperature.
   Wind and infall models overpredict the radio luminosities at small galaxy masses and have no particular 
   steepening of the run of the radio luminosities predicted at any galaxy mass. }
  % conclusions heading (optional), leave it empty if necessary 
   {Our analysis is consistent with radio AGN having the same characteristic, Eddington-scaled upper
   limit to jet powers in galaxies
   of all masses and with all galaxies having hydrostatic gas halos in phases when radio AGN are active. 
   We find no evidence for a change of the type of galaxy halo with galaxy mass.
   Galactic winds and quasi-spherical cosmological inflow phases cannot frequently happen at the same time 
   as powerful jet episodes unless the jet properties in these phases are significantly different from what we assumed
   in our model. }


   \maketitle
%
%-------------------------------------------------------------------

\section{Introduction}
Radio emission in galaxies is commonly attributed to activity of the supermassive black hole
as well as to processes related to the formation of stars 
\citep[for recent reviews see][]{Hardcastle15,Tadhunter16,Norris17,Krause18a}.
The LOFAR radio telescope is surveying the northern sky at low-frequency 
with unprecedented resolution and sensitivity \citep{Combea19}, giving us a much 
more comprehensive view of the radio emission of the local galaxy population. Of the 326,000
sources in the first data release of the LOFAR Two-metre Sky Survey \rev{\citep[LoTSS][]{Shimwea19}}, 
70\% have optical counterparts \citep{WilliamsWea19} including redshifts and absolute
magnitudes \citep{DuncanKea19}. 

\citet{Sabea19} cross-matched the LoTSS database
with the Sloan Digital Sky Survey and classified the sources as galaxies with or without
active galactic nuclei based on diagnostic emission line and two-colour plots, and 
the strength of the 4000~\AA~break and the luminosity in H$\alpha$ compared 
to the 150~MHz radio luminosity. This resulted in a local sample of 2121~galaxies with 
active galactic nuclei (AGN) and 
8494 star-forming galaxies. They find that galaxy mass is the main driver of 
radio-AGN formation and that above a stellar mass of $10^{11} M_\odot$, all galaxies
are switched on, whereas below $3 \times 10^{10} M_\odot$ less \rev{than} 10\% of the 
galaxies show even the faintest sign of a radio AGN. This is a significant
improvement of previous results, and a direct consequence of the more sensitive
LoTSS observations compared to previous surveys.

\rev{Separating contributions from AGN activity and
star formation to the total radio luminosity is, however, a} difficult issue \citep[e.g.,][]{Guerkea18}. 
It is even more complicated
by the fact that even luminous radio AGN frequently do not show any optical activity. 

The radio luminosity of an AGN depends on both, jet power and environment
\citep[e.g.,][]{KDA97,MK02,HK13,HK14,MacAlex14,TS15,EHK2016,Turnea18b}. The lack of 
radio AGN at lower galaxy masses might therefore either indicate that such galaxies 
have the capacity to produce radio jets to a much lesser extent, or that the gaseous
halos of the smaller galaxies differ significantly from the ones in bigger galaxies.
%We show below that neither of these assumptions is necessary.

Empirical studies provide hints that AGN may work differently in galaxies of different 
masses. For example, \citet{Bestea05} found that the expression of optical activity
is essentially independent of activity in the radio band, with lower mass galaxies
showing more optical activity and higher mass galaxies showing more radio-loud
AGN. Focussing on radio-loud AGN only, \citet{KHB08} showed that the probability
for them to have emission lines decreases with galaxy mass. 
This could, however, be due to the fact that at low galaxy mass, 
radio AGN with low Eddington ratio, which tend to have little line emission \citep{BH12}, 
are harder to detect.
On the other hand, the existence of scaling relations between black-hole mass 
and stellar velocity dispersion, bulge mass and total galaxy mass
\citep[e.g.,][]{Magea98,HR04,Guea09,ReinVol15,BentzMN18} suggests most galaxies 
have supermassive black holes, which grow in a similar way during phases of 
nuclear activity \citep[compare, e.g.,][]{Soltan82,MerlHeinz08,TucVol17}.
Radio AGN that also have strong emission lines tend to be associated
with more strongly star-forming hosts \citep{Hardcastlea13}, higher Eddington ratios
and lower stellar mass \citep{BH12}. Hence, the observational evidence suggests
that lower mass galaxies have, on average, higher Eddington-scaled accretion rates
on their central, supermassive black holes. In emission line AGN, the luminosity
in optical emission lines is broadly correlated to the radio luminosity 
\citep{MC93,KHB08}. Therefore, one should expect that the typical,
Eddington-scaled jet power in low-mass galaxies is, if anything, rather higher than in high-mass galaxies.

Galaxy halos, however, might be expected to be qualitatively different in galaxies with 
different masses: a hot halo can form via an accretion shock in galaxies with 
a mass of their dark matter halo $>10^{11} M_\odot$,
%\footnote{In this section, we rescale masses for galaxies 
%and dark matter halos given in the original papers to the mass in stars using the 
%halo occupation model of \citet{Mostea13}.}, 
whereas at lower masses, one expects 
that the accreting gas cools so fast that no shock forms \citep{BirnDek03}.
% Birnboim says 10^11 halo mass and Moster tells us this corresponds to 10^9 galaxy mass
This difference has lead to the concept that high mass galaxies accrete their gas mainly in the 
\emph{hot mode}, i.e. by first shock-heating it to the virial temperature of the dark matter
halo and subsequent gas cooling, whereas the lower mass galaxies accrete in the \emph{cold mode}
with cold gas being channeled into the galaxies along cosmic filaments \citep{DekBir06}. 
In a cosmological, hydrodynamic simulation, \citet{Kerea05} find a critical dark matter mass,
at a similar level of $3\times 10^{11} M_\odot$. We can use the halo occupation model of 
\citet{Mostea13} to derive a critical stellar mass of 1-3$\times 10^9 M_\odot$ where
the properties of the gas halos would be expected to change. 
However, this disagrees with the mass scale of $10^{11} M_\odot$ (stellar mass) now identified in the 
sensitive LOFAR measurements.

One might suspect that stellar feedback affects the properties of galaxy halos.
Two types of interstellar medium-halo interaction are known: \rev{{\em fountains}}, where the gas 
mainly circulates in the lower \rev{(i.e. closer to the galactic disc)} 
part of the halo and \rev{{\em galactic winds}} where gas appears at 
above escape speed and the outflow likely proceeds beyond the virial radius of the galaxy 
\citep[e.g.,][]{CC85,dAB04,dAB05,VCB05,DT08a,vGlea13,Gattoea15,HT17,KimOst18}.

In the fountain case, high entropy material accumulates at high altitudes and there is a 
smooth transition to a hydrostatic halo \citep[e.g.,][]{dAB04,dAB05}.
This is confirmed by studies of, e.g., the motion of HI clouds  
\citep[e.g.,][]{Mirabel81,KalbDed08,Marascea12}
and observations of the hot halo gas of the Milky Way in X-ray absorption
and emission \citep{GuptaAea12,GuptaAea17,Bregmea18}.

Simulations of galactic winds found that at the onset of the wind, any infalling or hydrostatic
halo is swept up by a shock wave, leaving the halo in a state 
of low density outflow with regions of turbulence and denser gas in filaments or 
on the walls of the outflow cone \citep[e.g.,][]{StSt00,Coopea08,DT08a,vGlea13,Ruszea17}.
Multi-wavelength observations of the different gas phases are generally consistent
with this structure \citep[e.g.,][]{VCB05,SH07,HT17}.

The transition between wind and fountain solution occurs at different galaxy masses in different
types of simulations, depending probably mainly on details of the feedback implementation.
%\footnote{We cite here only results based on grid codes, because of the strong dependence of the 
%properties of galactic winds on the gaseous properties of initial condition of the halo gas, 
%which is typically less well modelled in particle based methods.}. 
For example, \citet{DT08a}, using an effective equation of state for unresolved interstellar medium
and energy input from clustered supernovae,
find  a \rev{galactic} wind for their galaxies with circular velocity of 35~\kms~
(stellar mass $\approx 10^7M_\odot$), but a fountain for 75~\kms~ galaxies
 (stellar mass $\approx 10^9M_\odot$). 
 \citet{JacobSea18}, who also use an effective
 equation of state for unresolved interstellar medium and focus on
 cosmic rays as driver of feedback, find \rev{galactic winds below}
 160~\kms~(stellar mass $\approx 3\times 10^{10} M_\odot$),
 and a fountain solution at higher masses.
Galactic winds consistently form in the simulations, when a characteristic threshold
of star formation rate per unit area is exceeded \citep[e.g.,][]{vGlea13}. Hence,
any given galaxy might switch repeatedly between wind and fountain, depending on the 
current availability of fuel for star formation.
 
Direct observations of hot gaseous halos of galaxies in X-rays are rare \citep[for reviews see][]{Putmea12,TPW17}. 
For a few massive spirals and elliptical galaxies with stellar mass $\gtrsim10^{11}M_\odot$,
\citet{Bregmea18} report X-ray detections, and in some cases also density profiles.
The results are consistent with expectations for a gas halo at the virial temperature
close to hydrostatic equilibrium.
\citet{Strickea04b} show for a sample of disc galaxies with circular velocities between 100
and 244~\kms~(stellar masses: $\approx 10^{10}-10^{11} M_\odot$) that most of the luminosity of extraplanar
X-ray emission is likely related to superbubble blowout \citep[see also][]{Krausea14a}.
The total vertical extent of the X-ray halos, however, correlates with circular velocity and might therefore
also indicate a hydrostatic halo. Galaxies of all masses show multi-phase
gas in their halos \citep[e.g.][]{TPW17,Bordolea18,LanMo18}. Winds at or above escape 
velocity have been reported up to circular velocities of 300~\kms~\citep{HT17}.

Summarising, while trends of the state of gaseous halos (inflow, \rev{outflow or hydrostatic})
with galaxy mass are expected,
neither simulations nor observations currently provide a clear picture of these trends, or any
particular galaxy masses where transitions would occur.

However, the new LOFAR LoTSS survey has clearly identified such a critical galaxy mass at $\approx10^{11} M_\odot$.
Here we investigate, with simple models for the maximum luminosity expected
from the radio-AGN of a galaxy of a given mass, if this mass scale can be understood as a critical stellar mass
at which the properties of gaseous halos of galaxies change significantly. We first construct fiducial
gaseous halo models for the different situations described above (Sect.~\ref{s:halomodels}).
We then describe our models for the radio emission of AGN-jets in the given halos (Sect.~\ref{s:radioem}),
compare to the LoTSS radio emitting galaxies in Sect.~\ref{s:comp} and summarise our conclusions
in Sect.~\ref{s:conc}.

\section{Models for gaseous halos of galaxies}\label{s:halomodels}

\rev{In this section we discuss three simple models for gaseous halos of galaxies. Hydrostatic
halos are characterised by an overall equilibrium between gravity and pressure gradient.
Gas cooling, stellar and AGN feedback are unable to cause large perturbations to the overall
equilibrium, but produce convection (a galactic fountain) in the lower part of the halo close
to the stellar component, or buoyant bubbles.
Wind halos occur where feedback in the galaxy is strong enough to lead to a global gaseous
outflow from the galaxy beyond its virial radius. An inflow halo will occur, where pressure forces in the
halo and feedback from the galaxy are insufficient to balance the ram pressure of a global gas 
inflow into the galaxy. In reality, combinations of different halo types can occur in the same galaxy. For example,
a disc galaxy could have a global inflow in its equatorial region, when at the same time 
a starburst in its core drives an outflow in the polar directions. Since jets are a directed phenomenon, 
we consider here only galaxies with one type of gaseous halo and imply that this is the type 
of halo relevant for the direction into which the jets are emitted.}

\subsection{Hydrostatic halos}\label{ss:hydrostatic}
\begin{figure}
   \centering
   \includegraphics[width=\hsize]{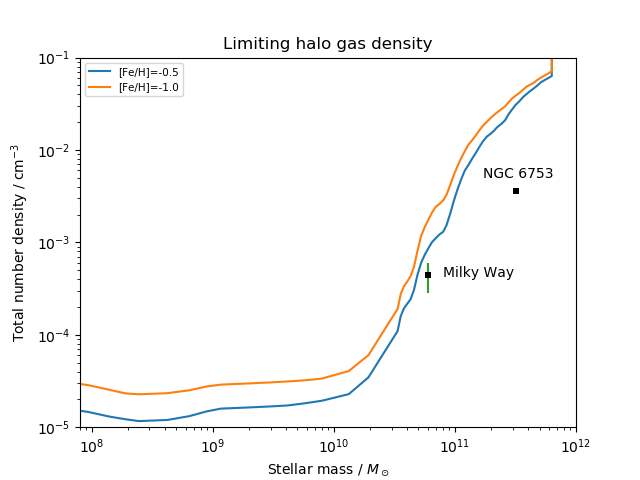}
      \caption{Maximum particle density of the hydrostatic halo as a function of stellar mass according to the model
      presented in Sect.~\ref{ss:hydrostatic} for two metallicities. For comparison, we show 
      measurements for the Milky Way 
      \citep[using the stellar mass estimate of $6\times10^{10} M_\odot$ from \citet{BR13} and \citet{TaylorCea16}]{GuptaAea17} 
      and the massive spiral galaxy
      \object{NGC 6753} \citep{Bogdea17}.
              }
         \label{f:maxden}
   \end{figure}

We construct a fiducial hydrostatic halo model from thermodynamic 
and cosmological constraints. Isothermal hydrostatic gas halos
in \citet{NFW97} dark matter halos were derived by \citet{Makinea98}. As they note, the resulting
profile is similar to the conventional $\beta$-profile, which we adopt in the following:
$\rho(r)=\rho_0 (1+(r/r_\mathrm{c})^2)^{-3\beta/2}$. Here $\rho(r)$ is the gas density profile
approximated as spherically symmetric and $r_\mathrm{c}$ is the core radius.
The \citet{Makinea98} gas profile is characterised by a core with a radius given by the scale
radius $R_\mathrm{s}$ of the dark matter halo, which is related to the virial radius
$R_\mathrm{vir}$ by the concentration $C=R_\mathrm{vir}/R_\mathrm{s}$. We use the
fitting formula from \citet{Klypea16} for redshift zero,
$C = 7.4 (M_\mathrm{vir}/(10^{12} M_\odot/h))^{-0.12}  (1+(M_\mathrm{vir}/M_0)^{0.4})$,
with $M_\mathrm{vir} =M_{200}= 200 \rho_\mathrm{crit} 4\pi R_\mathrm{vir}^3/3$ and 
$M_0=5.5\times10^{17} M_\odot/h$. We use the \citet{Planck16a} cosmology;
for redshift zero, $h =0.68$ and $\rho_\mathrm{crit}=8.62\times10^{-30}$g~cm$^{-3}$.
Massive spiral as well as elliptical galaxies are well fit by the $\beta$-model with $\beta\approx 0.5$
\citep{Bregmea18}
and we adopt this value. We note that this leads to a radio luminosity that declines with source
size for radio sources larger than the core radius. Since in the following we are only 
interested in the maximum luminosity a radio source can produce, the exact value of $\beta$ is 
not important as long as $\beta>0.37$, which is required for a radio luminosity
that declines with source size \citep{HK13,Yatea18}.
Again, since we are interested in the maximum radio luminosity of radio AGN for a given galaxy mass, 
we take the core density as the maximum density allowed by the requirement that the radiative
cooling time $t_\mathrm{c}$ exceeds the dynamical time. 
For a given dark matter halo, we define the dynamical
time as $t_\mathrm{dyn}=R_\mathrm{s}/v_\mathrm{vir}$ with the virial velocity $v_\mathrm{vir}=(GM_\mathrm{vir}/R_\mathrm{vir})^{1/2}$. We link virial masses to stellar masses $M_*$ by the halo occupation model of
\citet{Mostea13}, which we approximate as $\log_{10} M_* = 2.2\log_{10}M_\mathrm{vir} -15.4$
\rev{for $\log_{10} M_\mathrm{vir}<11.8$ and}  $\log_{10} M_* = 0.4\log_{10}M_\mathrm{vir} +5.8$, otherwise.
This yields dynamical times of the order of 100~Myr. Using the 
$\sbr{\rm Fe/H}=-0.5$ and $\sbr{\rm Fe/H}=-1.0$ \citep{Bogdea17,Bregmea18}
collisional ionisation equilibrium cooling functions $\Lambda(T)$ from \citet{SD93}, the maximum particle density
in a hydrostatic halo at the virial temperature is then given by
$n_\mathrm{max} = k_\mathrm{B}T_\mathrm{vir} / (\Lambda(T)\,t_\mathrm{c})$.
We plot the maximum particle density in Fig.~\ref{f:maxden} and compare it to measurements
for the Milky Way and a more massive spiral galaxy.  The strong increase between $M_*=10^{10}M_\odot$
and $M_*=10^{11}M_\odot$ mirrors the behaviour of the cooling function for the relevant virial temperature.
%The model yields reasonable upper limits up to several times $10^{11} M_\odot$. 
\rev{Towards} $10^{12} M_\odot$ the scale radii become 
very large, and the model would lead to a strong overestimate of the total gas mass. 
Therefore, we restrict the valid range of the model \rev{to $\log_{10} M_*/M_\odot \le 11.5$.} 
We adopt $\sbr{ \rm Fe/H}=-0.5$ \rev{as appropriate for the case of the Milky Way,
most galaxies where it has been measured \citep{Bregmea18} and the central galaxies of groups and clusters
\citep{BW10}. Assuming a metallicity of $\sbr{\rm Fe/H}=-1.0$ would increase the gas density
in the modelled halos by a factor of two or less. This would increase the predicted radio luminosities
by less than 50~\% \citep[compare, e.g.,][their Fig. 10]{MacAlex14},  
which is not important for the arguments presented here.}
%-------------------------------------------------------------
%                 Max density, hydrostatic halos
%-------------------------------------------------------------
%
%%%%%%%%%%%%%%%%%%%%%%%%%%%%%%%%%%%%%%%%%%%%%%%%%%%%
\subsection{Galactic-wind halo\rev{s}}\label{ss:gwh}
%%%%%%%%%%%%%%%%%%%%%%%%%%%%%%%%%%%%%%%%%%%%%%%%%%%%
Following the work of \citet{CC85}, we assume a core at almost constant density and pressure,
where the stellar mass and energy input take place,
and a free wind zone at constant outflow velocity with density declining as $r^{-2}$. 
The solution is fully determined by the energy and mass input rates,
$\dot M$ and $\dot E$, both linear in the star formation rate. The core temperature is proportional to
$\dot E / \dot M$
and therefore constant for all systems. 
The initial temperature could be as high as $\approx10$~keV 
\citep{CC85,SH07}. However, similar to the case of superbubbles
\citep{Krausea14a,RP14}, mixing of material from swept-up shells 
and remnant cloud cores likely reduces the temperature of the hot phase quickly. 
We therefore adopt a temperature of 3~keV which is approximately the lower limit
allowed by models for the nearby, probably best studied, wind galaxy
\object{M82} \citep{SH09}. The density in the model increases with star formation rate,
but for realistic conditions should be dominated by mixing as well.
Further following \citet{SH09}, we adopt
a core density of 0.6~cm$^{-3} M_*/M_{82}$, where $M_{82}$ is
the stellar mass of \object{M82}. 
The latter is between $10^9M_\odot$ \citep{FSea03} and $10^{10}M_\odot$
\citep[dynamical mass estimate]{Grecea12}. We adopt $M_{82}= 10^{10}M_\odot$,
\rev{because the lower value takes into account only the central parts of the galaxy.
Using the lower value would increase our core densities by a factor of ten, and the radio luminosities
for this model would consequently be higher by a factor of a few 
\citep[compare, e.g.,][their Fig. 10]{MacAlex14}. This would not change the
conclusions of the present analysis (compare below, e.g., Fig.~\ref{f:Sbtr-plot}).}
We use \rev{a} constant core radius of 500~pc, as for M82 \citep{Grecea12}, for all galaxies.
We note that the thermal pressure quickly declines outside the core, and hence
the isothermality assumed by our $\beta$-model ansatz is violated. We ignore this effect,
because we expect the maximum radio luminosity on the scale of the core radius.

%%%%%%%%%%%%%%%%%%%%%%%%%%%%%%%%%%%%%%%%%%%%%%%%%%%%
\subsection{Infall halo\rev{s}}\label{ss:infh}
%%%%%%%%%%%%%%%%%%%%%%%%%%%%%%%%%%%%%%%%%%%%%%%%%%%%
In the infalling halo picture, galaxies get their fuel for star formation from 
accretion of intergalactic gas. The accretion rate is therefore
given by the star formation rate. For this, we use the average star formation rate
of the main sequence enhanced  by one standard deviation from \citet{Belfea18}:
$-\log \dot M/M_\odot\,\mathrm{yr}^{-1} =$
$ \log \dot M_*/M_\odot\,\mathrm{yr}^{-1} = 0.73 \log(M_*/M_\odot) -6.94$.
While some of this gas might be clumpy, we get an upper limit on the gas density
by assuming spherically symmetric accretion. \rev{To model this, we consider either
a free fall or Bondi accretion into a \citet{NFW97} dark-matter halo.
The velocity close to the galaxy is in both cases given by  $v=-\sqrt{-2\Phi_\mathrm{NFW}}$, 
where the Navarro-Frenk-White potential is given by \citep{HNS07}:
\begin{equation}
\Phi_\mathrm{NFW} = -\frac{G M_\mathrm{vir}}{r} \frac{\log\left(1+r/R_\mathrm{s}\right)}{ \log(1+C) - C/(1+C) }\, .
\end{equation}
We then} get the halo
density from mass conservation, $\dot M = 4 \pi r^2 \rho v$. Since the infalling
medium is assumed to be cold, the relevant halo pressure is now the ram pressure 
of the infalling halo, $\rho v^2$. This leads to an almost constant velocity
in the relevant inner part of the halo, and hence a density and ram pressure distribution
as $r^{-2}$. This can be modelled by an isothermal beta profile with $\beta =2/3$.
We assume the infall continues down to 0.5~kpc, where we assume normal processes of star formation
to keep the density constant and the pressure balanced with the ram pressure of the infalling halo.

\section{Radio luminosities of jets in different halos }\label{s:radioem}
The early evolution of radio AGN is treated in \citet{Alex06} and \citet{MacAlex14}.
As can be seen, e.g., from Fig.~5 in \citet{MacAlex14}, the source luminosity increases
throughout the early evolution into the self-similar phase. It reaches its peak
approximately where the ambient density distribution starts to decline, i.e.,
near the core radius. In this phase, the radio lobes dominate the luminosity.
\rev{For an estimate of the radio luminosity, it is therefore appropriate to use a standard model
for the evolution of radio lobes, such as the one in \citet{Hardcastle18}. The model
requires that radio lobes have formed already, which first happens when 
the radio source has reached a certain size given, e.g., by \citet{Krausea12b}. 
After confirming that the radio lobes form earlier than our scale of interest for all cases,}
we therefore used the standard 
radio lobe models of \citet{Hardcastle18}.
Briefly, the model is based on 3D simulation results \citep{HK13,HK14,EHK2016},
distributes a constant fraction of the steadily supplied power $Q_0$, respectively,
to radio lobes and shocked ambient gas and advances the prolate spheroidal
outer shock surface according to the Rankine-Hugoniot shock jump conditions.
It assumes that the lobes consist of an electron-positron pair plasma (compare below)
and that the ratio between the energy in the magnetic field and the one in particles is~0.1.
We assumed an injection power law index for relativistic electrons of $q=-2.2$.
Adiabatic, synchrotron and inverse Compton losses at the cosmic microwave
background (redshift $z=0$) are taken into account. 
\citet{Sabea19} probe the jet power distribution for massive galaxies,
$10^{11}M_\odot<M_*<10^{12}M_\odot$, as a fraction of the Eddington
luminosity $L_\mathrm{Ed}$ of the supermassive black hole. They find a strong decline 
towards higher jet powers. Only 0.3\% of all galaxies have a radio AGN
with $Q_0>10^{-2}L_\mathrm{Ed}$. We therefore adopt this as upper limit
for radio sources in all galaxies. To estimate the black hole masses,
we adopt a constant mass fraction, $M_\mathrm{SMBH} =0.0005 M_*$
\citep{BentzMN18}.
%the scaling relation of \citet{BentzMN18}, 
%$M_\mathrm{SMBH} = 10^{8.4}M_\odot\, (M_*/10^{11}M_\odot)^{1.84}$,
%multiplied by a factor of three to take into account the scatter, since we
%are interested in the maximum plausible luminosity for a given galaxy mass.

\begin{figure}
   \centering
    \includegraphics[width=\hsize]{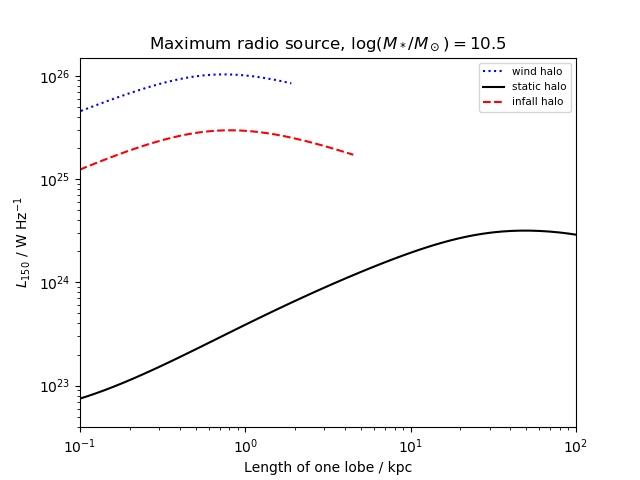}
     \caption{Example radio luminosity at 150~MHz versus lobe length plots for hydrostatic,
      galactic wind and infall halos for a galaxy of stellar mass $M_*=3 \times10^{10}M_\odot$
      and the maximum plausible jet power in our model of $2\times10^{36}$~W.
      See Sect.~\ref{s:halomodels} for details of the gaseous halo models and Sect.~\ref{s:radioem}
      for details of the radio source model.
              }
         \label{f:explPD}
   \end{figure}

We show a plot of the resulting radio luminosities at 150~MHz over lobe length
in Fig.~\ref{f:explPD} for an intermediate mass galaxy. 
Due to our assumption that wind and infall halos
have a small core region of only 0.5~kpc, the radio luminosity also peaks 
on this scale.
The relevant length scale for the hydrostatic halo is 
the scale radius of the dark-matter halo. Consequently, the peak of the
radio luminosity is reached at a scale of tens of kpc. The peak luminosities
for the wind and infall models are much higher than for the hydrostatic-halo case.
This is because, at the galaxy mass of the chosen example, $M_*=3\times 10^{10}M_\odot$,
cooling still severely restricts the density in a hydrostatic halo.
The luminosities in the wind and infall models would increase even further, if 
the core radius was assumed to be greater than 0.5~kpc. 
\rev{The radio luminosity in our models scales directly with the kinetic jet power,
which is linearly coupled to the black hole mass. Varying the black hole mass to galaxy
mass ratio would therefore move all models along the vertical axis in Figs.~\ref{f:explPD}~and~\ref{f:Sbtr-plot}. }

%-------------------------------------------------------------
%                 Example P-D tracks
%-------------------------------------------------------------

%%%%%%%%%%%%%%%%%%%%%%%%%%%%%%%%%%%%%%%%%%%%%%%%%%%%
\section{Comparison to the LoTSS sample}\label{s:comp}
\begin{figure}
   \centering
   \includegraphics[width=\hsize]{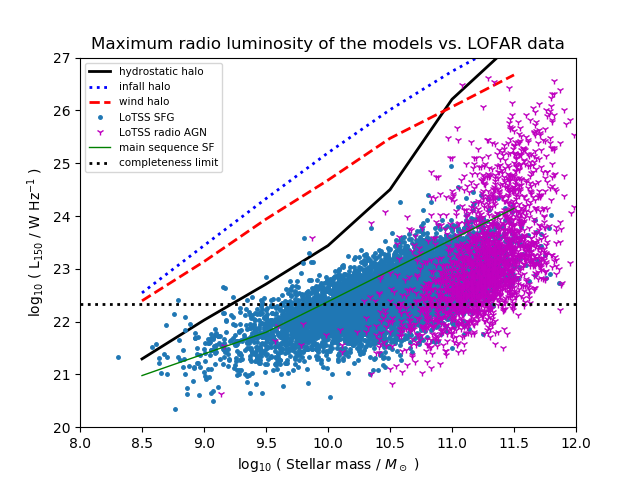}
      \caption{Maximum 150~MHz luminosity of radio AGN for a given galaxy mass for 
      three different models of the gaseous halo of the galaxies (hydrostatic halo, 
      infall halo, wind halo). Data points are from the 
      LoTSS sample \citep{Sabea19}, plotted separately for star forming galaxies (SFG)
      and radio AGN. The radio luminosity expected from star formation for galaxies
      on the main sequence of star formation at redshift zero is indicated as solid thin green line
      \citep[Sect.~\ref{ss:infh}, eq.~(3) in][]{Guerkea18}.
      The dotted black line denotes the completeness limit,
      i.e., the median limiting luminosity at the maximum redshift of the sample.
      For the hydrostatic halo model, radio AGN
      with the same Eddington-scaled jet power distribution could be hosted by galaxies of all masses.
      The lower densities in the halos of lower mass galaxies would, however, limit their
      radio luminosities.
      Radio AGN need to be intrinsically less powerful or otherwise different in low mass galaxies that
      are in galactic wind or (any) quasi-spherical infall phases in order not to exceed
      the LOFAR radio luminosity constraints.
              }
         \label{f:Sbtr-plot}
   \end{figure}

%%%%%%%%%%%%%%%%%%%%%%%%%%%%%%%%%%%%%%%%%%%%%%%%%%%%
We compare the maximum of the predicted radio luminosities against
galaxy mass for all halo models to the LoTSS measurements in Fig.~\ref{f:Sbtr-plot}.
The radio sources in the hydrostatic halos approximate \rev{well} the upper envelope
of the data points. It is well known that the models reproduce radio luminosities
of radio sources in large dark-matter halos well \citep{HK13,Hardcastle18}, it is 
an interesting feature of the present model that the radio luminosities of smaller
galaxies are also not overpredicted by a large factor. 
This means that, if all galaxies have hydrostatic gas halos,
radio AGN in low-mass galaxies could have the same jet-power distribution (in terms of the Eddington
luminosity of the supermassive black hole) as the ones in high-mass galaxies,
but the observed luminosities would be much lower. The low luminosities
are due to the lower halo gas densities, which are a direct consequence 
of the higher cooling rates at lower halo temperature,
which via the virial temperature is a function of the galaxy mass (compare Fig.~\ref{f:maxden}).
Of course the radio luminosity in any given low-mass galaxy might still be entirely due
to star formation. The models only tell us that, if the galaxy developed a radio AGN,
its radio luminosity would not exceed a certain luminosity. The models therefore need 
to be compared to the upper envelope of the observed distribution.

Radio sources in wind galaxies or in halos dominated by spherical infall would have a
luminosity far greater than observed. This could be interpreted as galaxies 
not having strong radio AGN in such phases, or that such conditions are rare.
This would agree with findings in the literature that starburst and AGN phases are sequential
rather than simultaneous \citep[e.g.,][]{Krause2005b,Schawea07,Schartea10,Shabea12a}.
The predicted luminosities become very similar at high galaxy masses. \rev{At those masses,}
all halo models are consistent with the radio observations.

An interesting caveat here is the type of the radio source. It is possible that jets
in more strongly star-forming systems suffer much more entrainment of proton-rich
interstellar medium, and therefore develop more \citet{FR74} class~I-like radio sources \citep{CIH18}.
In this case, the same jet power would be distributed also to additional protons, and thus
the expected radio luminosity would be lower than in our model, possibly to a degree
that could reconcile the prediction for wind and infall halos with the observations.

The spatial resolution of the LoTSS survey corresponds to 28~kpc at the redshift limit of $z=0.3$
\citep[6~arcsec resolution,][]{Sabea19}. 
Radio AGN in infall and wind halos would therefore mostly be unresolved. The maximum radio luminosity
in hydrostatic halos occurs around 20 (40, 100)~kpc for $10^9 (10^{10}, 10^{11}) M_\odot$ galaxies.
Radio AGN would therefore also frequently be unresolved in less massive galaxies in our model, if they had 
a hydrostatic halo. This agrees with the findings of \citet{Shabala18} that the fraction of 
unresolved radio AGN decreases with increasing galaxy mass. This prediction could be tested
with LOFAR observations that include international baselines \citep[$<1$~arcsec resolution,][]{Ramirea18}, which are,
however, not yet available for large samples of galaxies.

The maximum radio luminosity we predict in our models should not be regarded as a strict upper limit
on the measured LOFAR 150~MHz luminosities. We estimate the measurement uncertainties 
for the luminosities due to uncertainties of the flux calibration scale  
to about 0.3~dex or more \citep[compare][]{Hardcastlea19}.
Also,  jet powers are known to sometimes exceed our choice of 1\% of the Eddington luminosity
\citep{Sabea19}. For example \citet{Turnea18b} estimate jet powers around $10^{47}$~erg/s
for several 3C radio sources in galaxies with $11.5<\log(M_*/M_\odot)<12$. The jet power
in these sources therefore likely exceeds 10\% of the Eddington luminosity. Further,
the black-hole-mass scaling relation has a scatter of about a factor of three. Taken together,
individual sources could still be one or two dex more luminous than predicted by our model.

The high jet power phases might, however, be rare, possibly linked to galaxy merging
\citep{Ramea12,Shabea12a,Tadhea14,Krausea19a}. The \citet{Sabea19} sample is local 
(redshift $z\le0.3$) and therefore contains relatively few recent mergers,
especially for the lower galaxy masses \citep{Hopkea10}. 

For the general distribution of galaxies, however, our assumptions reasonably cover
the top end of the relevant distributions: in \citet{Sabea19}, much less than 1\% of the
radio sources exceed our jet power limit and the halo properties we assume take 
account of the scatter. For example, Fig.~\ref{f:maxden} demonstrates that observations fall 
comfortably below our limiting halo gas density.

%
%-------------------------------------------------------------
%                 Sabater plot
%-------------------------------------------------------------
%

%%%%%%%%%%%%%%%%%%%%%%%%%%%%%%%%%%%%%%%%%%%%%%%%%%%%
\section{Conclusions}\label{s:conc}
%%%%%%%%%%%%%%%%%%%%%%%%%%%%%%%%%%%%%%%%%%%%%%%%%%%%
We use an established model for the radio luminosity of radio AGN
to estimate the maximum AGN-related LOFAR 150~MHz luminosities of 
galaxies. As input to our model, we assume a jet power of 1\% of 
the Eddington luminosity with supermassive black hole masses following the
observed scaling relation and models for the gaseous halos of galaxies based
on theoretical constraints and observations. We consider hydrostatic,
galactic wind and infall halos. Our aim is to explain the marked rise of radio luminosities
at stellar masses around $10^{11} M_\odot$.
Our main findings are:
   \begin{enumerate}
       \item The shape of the cooling function translates into higher hydrostatic halo masses, and hence 
      a marked increase in radio luminosity around $10^{11} M_\odot$.
     \item Assuming hydrostatic gas halos in all galaxies and the jet power (as fraction
      of the Eddington luminosity) distribution seen in massive galaxies to hold for all galaxies
      we find an upper envelope for the radio luminosities that matches the data well.
      \item Disproportionately lower radio luminosities in low-mass galaxies can therefore
      not be used as an argument that radio-AGN are absent in such galaxies.
      \item A model where low-mass galaxies are dominated by winds or infall and higher 
      mass galaxies by hydrostatic gas halos is not supported by the data.
      \rev{Models expect the presence of hydrostatic halos in galaxies with masses 
      $M_*\gtrsim10^9M_\odot$ \citep{BirnDek03,Kerea05}.
      This corresponds to the mass range covered by the LOTSS sample. Our findings are hence consistent
      with these predictions.}
      \item We would predict higher radio luminosities than observed for low-mass galaxies, if their
      jets were frequently launched in phases of smooth, quasi-spherical inflow
      or galactic winds, unless jets in such environments suffer a lot of entrainment
      of proton-rich gas. This is consistent with the idea that galactic winds and 
      AGN-jet outbursts do not happen at the same time.
      \item Our results are also consistent with scenarios in which radio AGN occur
      much more rarely in lower mass galaxies, such that their radio luminosities are always
      completely dominated by star formation. We regard, however, the alternative, namely 
      that AGN have similar jet power distributions at all galaxy masses, which is as well consistent with the data, 
      as the simpler explanation.
   \end{enumerate}

\begin{acknowledgements}
 We gratefully acknowledge the provision of the LOFAR radio data in electronic form 
 by Jose Sabater Montes. We thank the anonymous referee for a very useful report.
\end{acknowledgements}

%-------------------------------------------------------------------
%
   \bibliographystyle{aa} % style aa.bst
   \bibliography{/Users/mghkrause/texinput/references} % your references Yourfile.bib
%
% - join the .bib files when you upload your source files
%-------------------------------------------------------------------
\end{document}